\documentclass[doublecol]{epl2}
\usepackage{amsmath}
\usepackage{bm}
\usepackage{amssymb}

\title{Quantum oscillations and upper critical magnetic field of the iron-based superconductor FeSe}
\shorttitle{Quantum oscillations and upper critical magnetic field of FeSe} %Insert here a short version of the title if it exceeds 70 characters

\author{Alain Audouard\inst{1} \and Fabienne~Duc\inst{1}  \and Lo\"{\i}c Drigo\inst{1} \and Pierre Toulemonde\inst{2} \and Sandra Karlsson\inst{2} \and Pierre Strobel \inst{2}\and Andr\'{e} Sulpice\inst{3} }
\shortauthor{A. Audouard \etal}

\institute{
  \inst{1} Laboratoire National des Champs Magn\'{e}tiques
Intenses (UPR 3228 CNRS, INSA, UJF, UPS) 143 avenue de Rangueil,
F-31400 Toulouse, France.\\
  \inst{2} Univ. Grenoble Alpes, Institut N\'{E}EL, F-38000 Grenoble, France.  CNRS, Institut N\'{E}EL, F-38000 Grenoble, France.\\
  \inst{3} Univ. Grenoble Alpes, CRETA, F-38000 Grenoble, France. CNRS, CRETA, F-38000 Grenoble, France.\\
  }


\abstract{
Shubnikov-de Haas (SdH) oscillations and upper critical magnetic field ($H_{c2}$) of the iron-based superconductor FeSe ($T_c$ = 8.6 K) have been studied by tunnel diode oscillator-based measurements in magnetic fields of up to 55 T and temperatures down to 1.6 K. Several Fourier components enter the SdH oscillations spectrum with frequencies definitely smaller than predicted by band structure calculations indicating band renormalization and reconstruction of the Fermi surface at low temperature, in line with previous ARPES data. The Werthamer-Helfand-Hohenberg model accounts for the temperature dependence of $H_{c2}$ for magnetic field applied both parallel (\textbf{H} $\|$ $ab$) and perpendicular (\textbf{H} $\|$ $c$) to the iron conducting plane, suggesting that one band mainly controls the superconducting properties in magnetic fields despite the multiband nature of the Fermi surface. Whereas Pauli pair breaking is negligible for \textbf{H} $\|$ $c$, a Pauli paramagnetic contribution is evidenced for \textbf{H} $\|$ $ab$ with Maki parameter $\alpha$ = 2.1, corresponding to Pauli field $H_{P}$ = 36.5 T.}

\begin{document}

\maketitle

\section{\label{sec:Intro}Introduction}

The discovery of iron-based superconductors \cite{Ka08,Re08,Di08,Ye08} has reactivated the questioning about the interplay, either competition or cooperation, between magnetism and superconductivity in correlated electron systems. Indeed, as the temperature decreases, parent phases of iron-pnictide superconductors, such as BaFe$_2$As$_2$, undergo a tetragonal-orthorhombic transition closely linked to the condensation of a spin-density wave (SDW), i.e. to the development of an antiferromagnetic long range order. In contrast, concomitant decrease of the N\'{e}el temperature and superconducting critical temperature rise is observed on doping in both Ba(Fe$_{1-x}$Co$_x$)$_2$As$_2$ (electron doped) or Ba$_{1-x}$K$_x$Fe$_{2}$As$_2$ (hole doped) \cite{Ru09,Bo14}, suggesting competition between SDW and superconductivity \cite{Zh11}.

Even though no clear quantitative consensus has been reached yet for the iron-chalcogenide FeTe$_{1-x}$Se$_x$ phase diagram, it can be stated that superconductivity emerges as $x$ increases (from $x$ in the range 0 $\sim$ 0.3) from a SDW state, similarly to the case of pnictide compounds \cite{Do13} even though Te and Se are isovalent. However, at variance with parent phases of iron pnictide superconductors, no long range magnetic order has been detected in the FeSe superconductor although the tetragonal-orthorhombic phase transition is observed at $\sim$ 90 K. Nevertheless, according to NMR data, antiferromagnetic spin fluctuations are strongly enhanced at temperatures close to T$_c$ \cite{Im09} which suggests that spin fluctuations may nonetheless play an important role in superconductivity. In line with this statement, itinerant SDW instability have been proposed \cite{Su08}.

Within this picture, nesting properties, hence Fermi surface (FS) topology, may play a major role for superconductivity of iron-based superconductors \cite{Ca11}. According to band structure calculations based on density functional theory, the FS of FeSe is composed of 2 concentric quasi-two dimensional electron and 3 hole tubes, located at the corner and center of the first Brillouin zone, respectively, with their axis parallel to the $c^*$ direction \cite{Su08,Ku10,Ma13,Te14}. Although these tubes are significantly corrugated, imperfect nesting can be considered \cite{Su08}, as in the case of iron-pnictide superconductors. However, strong discrepancy between band structure calculations and ARPES data has been reported \cite{Ma13}. Namely, much smaller tube areas have been observed and interpreted on the basis of a strong band renormalization and Fermi energy shift. In that respect, the tetragonal-orthorhombic transition at 90 K, connected to a nematic state with an orbital character, plays a significant role \cite{Na14,Sh14}. Nevertheless, it can be remarked that SDW with imperfect nesting would also lead to small tube area due to FS reconstruction.

Quantum oscillations study is a powerful tool to obtain information on the FS \cite{Ca11,Co13}.  As an example, the de Haas-van Alphen oscillations spectrum of LaFe$_2$P$_2$, which is a non-superconducting parent of iron-pnictide superconductors, is in agreement with band structure calculations \cite{Bl14}. This feature, which demonstrates the absence of any nesting of the FS at low temperature, is in conflict with the above mentioned picture of competition between SDW and superconductivity. Oppositely, Shubnikov-de Haas oscillations with frequencies in the range 60 T to 670 T, corresponding to orbits area from 0.2 to 2.3 $\%$ of the first Brillouin zone (FBZ) area, much smaller than predicted by band structure calculations, have been very recently observed in FeSe \cite{Te14}. Even lower frequencies, in the range 45-230 T had been previously reported for thin non-superconducting FeTe$_{1-x}$Se$_x$ crystals\cite{Ok13}, this latter compound having similar FS topology.

Besides quantum oscillations, temperature dependence of the upper critical magnetic field $H_{c2}$ may provide information regarding both the superconducting gap topology and the either single or multiband nature of the superconductivity which are both related to the FS topology \cite{Zh11,Co13}. As it is the case for e.g. FeTe$_{0.5}$Se$_{0.5}$, discrepancies can be observed within the few $H_{c2}$ measurements reported for FeSe \cite{Le12,Ve13,Te14}.

The aim of this article is to report on Shubnikov-de Haas oscillations to get more insight on the Fourier spectrum, hence the FS topology and band renormalization, and upper critical magnetic field measured by contactless tunnel diode oscillator technique (TDO) on high quality FeSe single crystals.

\section{\label{sec:Experimental}Experimental}

Studied single crystals have been grown using the chemical vapor transport method in sealed quartz tube, starting from Fe and Se powders (with a 1.1:1 molar ratio) in an eutectic KCl+AlCl$_3$ chlorides mixture as detailed in Ref.~\cite{Ka14}. The temperature profile used was inspired from the previous work of Chareev et al. \cite{Ch13} and B\"{o}hmer et al.\cite{Bo13} with a gradient temperature of 120$^{\circ}$C maintained during 6 weeks between the hot zone (440$^{\circ}$C) and the cold zone (320$^{\circ}$C) of the furnace. The average composition of the obtained crystals was determined to be Fe$_{1.02(1)}$Se by EDX micro-analysis of the surface of different crystals in a SEM.

As reported in Ref.~\cite{Dr10}, the device for radio frequency measurements is a LC-tank circuit powered by a
TDO biased in the negative resistance
region of the current-voltage characteristic. This device is connected to a pair of compensated coils made with copper wire (40 $\mu$m in
diameter). Each of these coils is wound around a Kapton tube of 2 mm in diameter. The studied crystal, which is a platelet with dimensions of roughly 1.4 $\times$ 1.4 $\times$ 0.04 mm$^3$, is placed at the center of one of them. The fundamental resonant frequency $f_0$ of the whole set is $\sim$
24 MHz. This signal is amplified, mixed with a frequency $f$ about 1 MHz below the fundamental frequency and demodulated. Resultant frequency $\Delta f$ = $f - f_0$ has been measured in zero-field and in pulsed magnetic fields of up to 55 T with a pulse decay duration of 0.32 s in the temperature range 1.5 K to
9 K. It has been checked that the data collected during the raising and the falling part of the pulse coincide, indicating that  no discernible temperature change occurred during the field sweep. It must be kept in mind that the TDO frequency is sensitive to the resistivity, yielding Shubnikov-de Haas oscillations\cite{Dr10}, and the London penetration depth\cite{Pr11}, in the normal and superconducting states, respectively.

\section{\label{sec:Results}Results and discussion}

%-----------------------------RT-------------------------
\begin{figure}
\centering
\includegraphics[width=0.95\columnwidth, clip,angle=0]{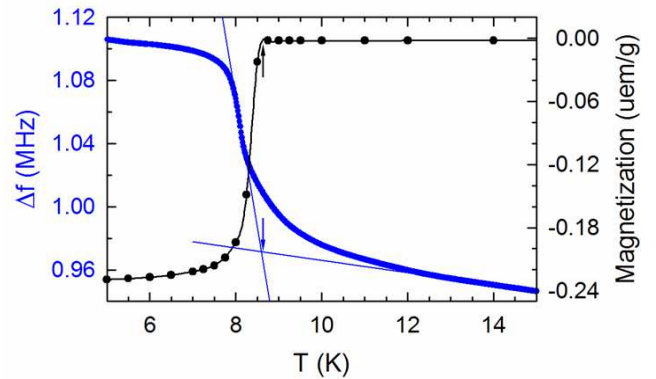}
\caption{(Color online)  Temperature dependence of the zero-field TDO frequency and zero-field cooled magnetization. Construction lines for the determination of $T_c$ are displayed. }
\label{fig:RT}
\end{figure}
%-----------------------------------------------------------------

As previously reported for other iron-based superconductors \cite{Ga11,Au14}, the zero-field TDO frequency displayed in Fig.~\ref{fig:RT} evidences a large increase as the temperature decreases linked to the decrease of the London penetration depth \cite{Pr11}. The onset of this frequency rise coincides with the onset of the zero-field cooled magnetization decrease and is therefore regarded as the superconducting transition temperature ($T_c$ = 8.6 $\pm$ 0.1 K).

\subsection{Upper critical magnetic field}

%-----------------------------transitions-------------------------
\begin{figure} [h]
\centering
\includegraphics[width=0.95\columnwidth, clip,angle=0]{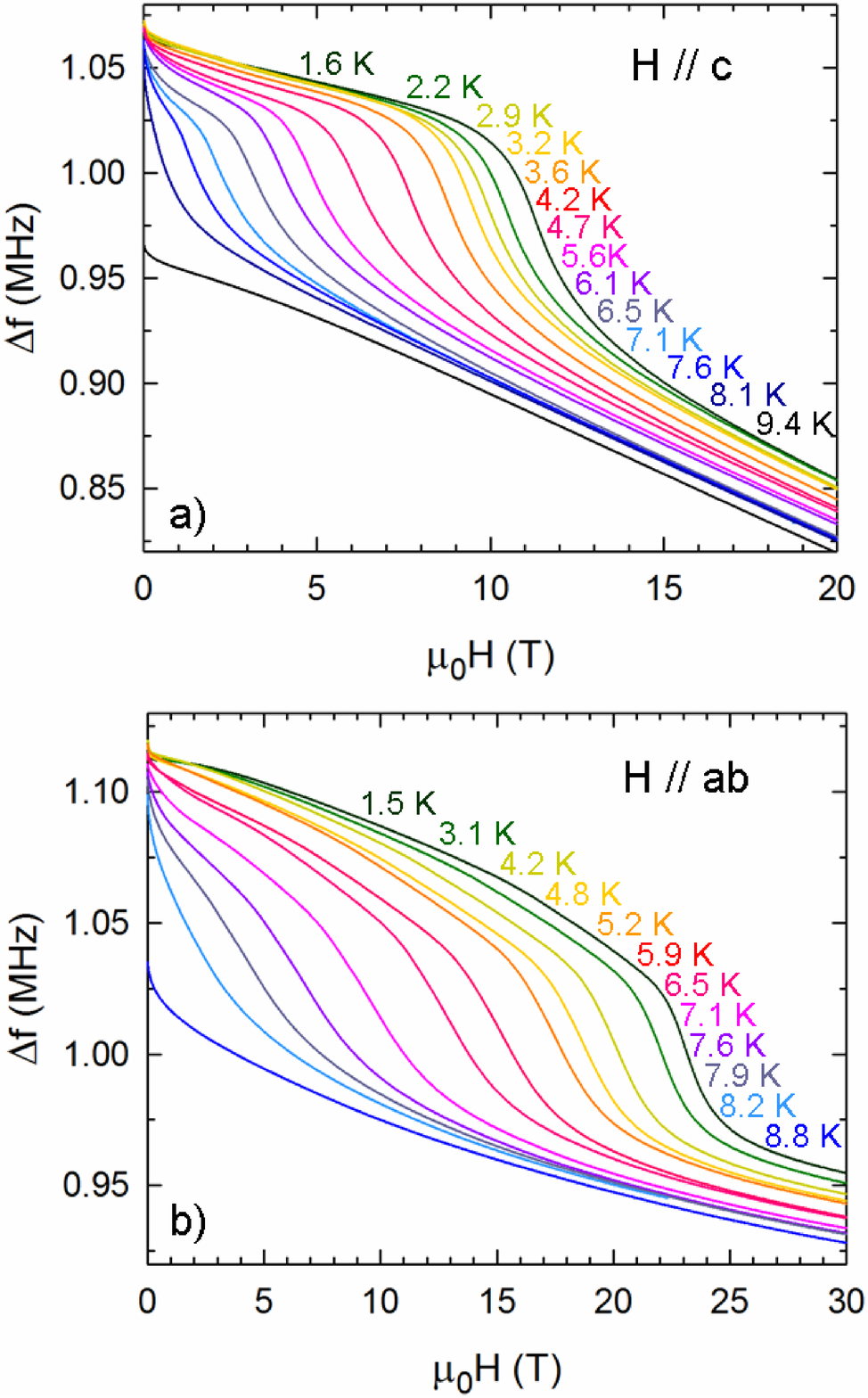}
\caption{(Color online) Field dependence of the TDO frequency for \textbf{H} perpendicular (a) and parallel (b) to the iron conducting plane, respectively. }
\label{fig:f(B)}
\end{figure}
%-----------------------------------------------------------------

%-----------------------------Hc2---------------------------------
\begin{figure}
\centering
\includegraphics[width=0.95\columnwidth, clip,angle=0]{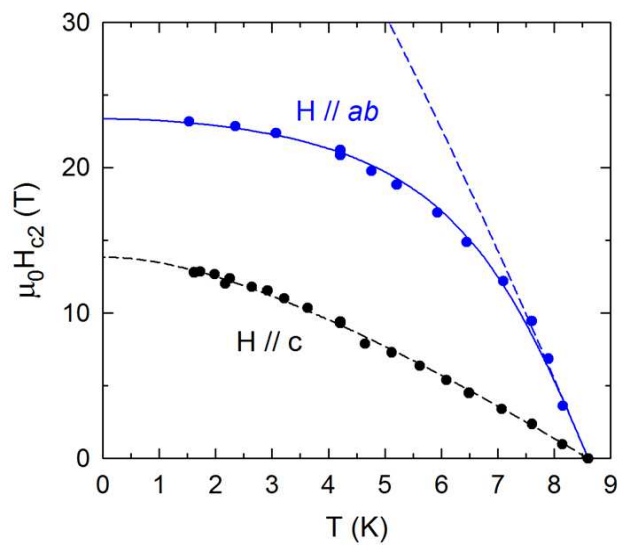}
\caption{(Color online) Temperature dependence of the upper critical magnetic field for \textbf{H} $\|$ $c$ (black symbols) and \textbf{H} $\|$ $ab$ (blue symbols). Dashed lines are the best fits of the WHH model to the data in the whole temperature range and the high temperature range for \textbf{H} $\|$ $c$ and  \textbf{H} $\|$ $ab$, respectively. Solid line is the best fit to the data for  \textbf{H} $\|$ $ab$ including a Pauli contribution ($\mu_0H_{P}$ = 36.5 T).}
\label{fig:Hc2}
\end{figure}
%-----------------------------------------------------------------

Field-dependent TDO frequency is displayed in Fig.~\ref{fig:f(B)} for magnetic field applied perpendicular (\textbf{H} $\|$ $c$) and parallel (\textbf{H} $\|$ $ab$) to the conducting $ab$ plane, at various temperatures. Marked transitions are observed for the two considered field directions. These data allow to reliably determine the temperature dependence of the upper critical magnetic field $H_{c2}$, reported in Fig.~\ref{fig:Hc2}, in a similar way to the $T_c$ determination (see Fig.~\ref{fig:RT} and Ref.~\cite{Au14}). These data yield  $dH_{c2}^{ab}/dT\mid_{T=T_c}$ = -9.2 T/K and  $dH_{c2}^{c}/dT\mid_{T=T_c}$ = -2.3 T/K. Despite these values are by a factor of 1.4 higher than those deduced from resistivity data  of Terashima et al. \cite{Te14}, they yield an anisotropy parameter $\gamma$ = 4.0 at T = T$_c$, in agreement with the data of Ref.~\cite{Te14}. This value is higher than the reported value of Braithwaite et al. ($\gamma$ = 1.4) \cite{Br09} and  Vedeneev et al. ($\gamma \sim$  1.5-2) \cite{Ve13}. This discrepancy could be attributed to better crystal quality in the two former cases. In that respect, it has been evidenced that e.g. columnar defects induced by heavy ion irradiation decrease $\gamma$ \cite{Ye13}. Otherwise, the measured $\gamma$ value is also higher than the anisotropy parameter of FeTe$_{0.5}$Se$_{0.5}$ crystals, determined with the same measurement technique ($\gamma$ = 1.6) \cite{Au14}, suggesting a stronger two-dimensional character in the former case.

Dashed lines in Fig.~\ref{fig:Hc2} are the best fits of the Werthamer, Helfand, and Hohenberg (WHH) model\cite{WHH} to the data, assuming weak coupling \cite{Zh11}. In this framework, the temperature-dependent upper critical field is given by ln(1/t) = $\psi(1/2+h/2t)-\psi(1/2)$ where $\psi$ is the digamma function, $t=T/T_c$ and $h=4\mu_0H_{c2}/[\pi^2(-dH_{c2}/dt)\mid_{t=1}]$. Orbital fields deduced within this framework  ($\mu_0H_{c2}^c(0)$ = -0.693$T_c dH_{c2}/dT\mid_{T=T_c})$ are $\mu_0H_{c2}^c(0)$ = 14 T and   $\mu_0H_{c2}^{ab}(0)$ = 55 T. These values are lower than those deduced from TDO data of FeTe$_{0.5}$Se$_{0.5}$ ($\mu_0H_{c2}^c(0)$ = 49 T and   $\mu_0H_{c2}^{ab}(0)$ = 78 T)\cite{Au14}. Hence, the deduced coherence lengths ($\xi_{c}$ =$\sqrt{(\phi_0 H_{c2}^c(0))/(2\pi)}/H_{c2}^{ab}(0)$ =  1.2 nm and $\xi_{ab}$ =$\sqrt{\phi_0/2\pi H_{c2}^c(0)}$ = 4.9 nm), which are close to the values deduced from resistivity measurements \cite{Te14}, are larger to that deduced from TDO data  for FeTe$_{0.5}$Se$_{0.5}$. Even smaller coherence lengths are deduced from specific heat data of FeTe$_{0.5}$Se$_{0.5}$ \cite{Kl10} suggesting larger effective mass, hence stronger renormalization for FeTe$_{0.5}$Se$_{0.5}$.
The orbital field for \textbf{H} $\|$ $c$  yields, according to the Clogston formula \cite{Cl62}, a superconducting gap  $\Delta$ = 1.1 $meV$, i. e. 2$\Delta$=3.1$k_BT_c$ which is very close to the weak coupling BCS value (2$\Delta$=3.5$k_BT_c$). This gap value is in agreement with muon-spin rotation \cite{Kh10} and specific heat data \cite{Li11}. However, larger gaps are deduced from  ARPES data \cite{Ma13} which suggests strong coupling instead.

While orbital effects account alone for the temperature dependence of $H_{c2}$ for \textbf{H} $\|$ $c$, Pauli pair breaking contribution must be included for \textbf{H} $\|$ $ab$. In this case, the orbital critical field is reduced as $\mu_0H_P=\mu_0H_{c2}^{orb}/\sqrt{1+\alpha^2}$ where the Maki parameter is given by  $\alpha = \sqrt{2}H_{c2}^{orb}/H_P$. A very good agreement with experimental data is obtained with a Pauli field $\mu_0H_P$ = 37 T, $i. e.$ $\alpha$ = 2.1 (see solid line in Fig.~\ref{fig:Hc2}) yielding $\mu_0H_{c2}(0)$ = 23 T. Due to the contribution of the Pauli effect for \textbf{H} parallel to the $ab$ plane, the anisotropy parameter decreases down to $\gamma$ = 1.7 as the temperature goes to zero. Nevertheless, the still unexplained anisotropy inversion (in which $H_{c2}^{ab}$ is lower than $H_{c2}^c$ below $\sim$ 4 K),  reported for FeTe$_{0.5}$Se$_{0.5}$ \cite{Se10,Br10,Au14} is not observed for FeSe. Noticeably, multiband superconductivity observed in muon-spin rotation data \cite{Kh10} and inferred from the almost linear temperature dependence of $H_{c2}^{c}$ and $H_{c2}^{ab}$ upturn at low temperature reported by Terashima et al.\cite{Te14} is not detected in the temperature dependence of $H_{c2}$. This result suggests that despite the reported multiband nature of the Fermi surface, superconducting properties are dominated by one band, possibly due to strongly different diffusivities of the various bands \cite{Gu03} as discussed for 1111 and 122 arsenides by Lei et al. \cite{Le10}. The different behaviour reported in Ref.~\cite{Te14} could be ascribed to the influence of vortices dynamics on conductivity data below $T_c$.
\\

\subsection{Shubnikov-de Haas oscillations}

%-----------------------------d_TF--------------------------------
\begin{figure}
\centering
\includegraphics[width=0.95\columnwidth, clip,angle=0]{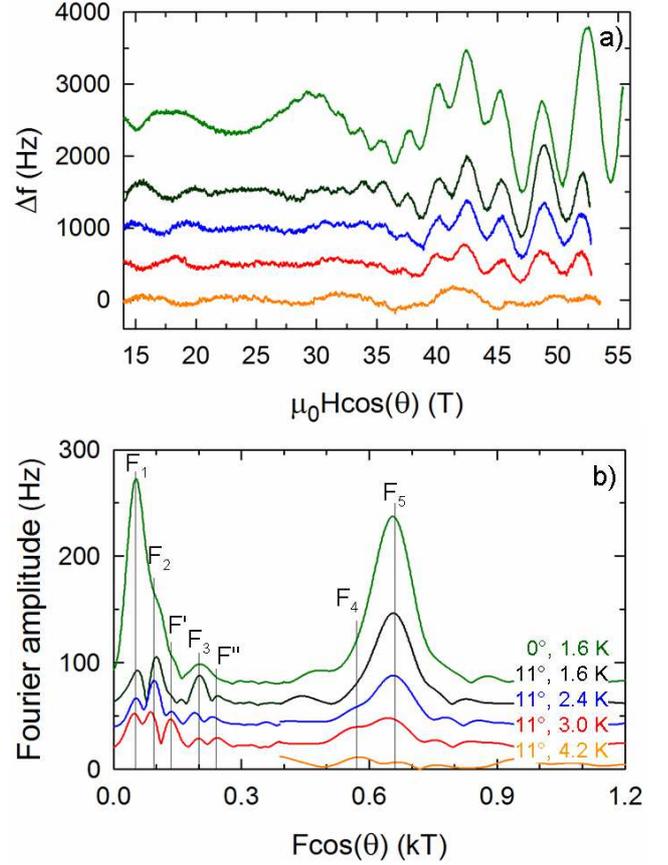}
\caption{(Color online) (a) Field-dependent oscillatory part of TDO signal at various temperatures. (b) corresponding Fourier analysis in the field range 15-55 T and 25-55 T below and above 0.4 kT, respectively. Salient frequency peaks are marked as thin lines (see text). The temperature and angle $\theta$ between the field direction and the normal to the conducting plane is indicated in (b).}
\label{fig:d_TF}
\end{figure}
%-----------------------------------------------------------------

%-----------------------------Dingle------------------------------
\begin{figure}
\centering
\includegraphics[width=0.95\columnwidth, clip,angle=0]{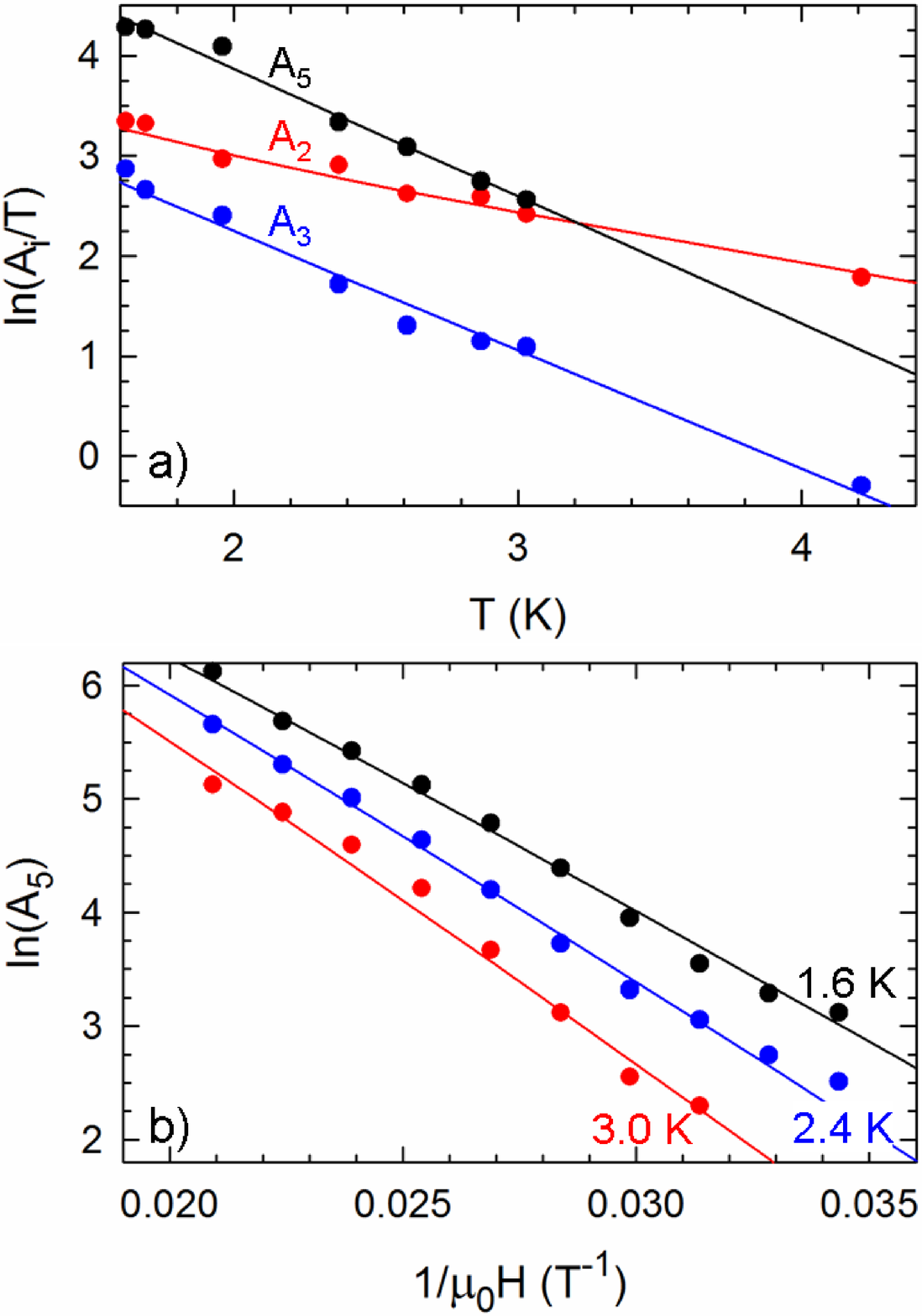}
\caption{(Color online) (a) Temperature dependence of the Fourier amplitude of components with frequency $F_2$, $F_3$ and $F_5$. (b) Field dependence of the Fourier amplitude $A_5$ at various temperatures. Solid lines are best fits of the Lifshitz-Kosevich model yielding effective masses ($m_2$ = 0.75, $m_3$ = 2.0 and $m_5$ = 3.2, in $m_e$ units) and Dingle temperature ($T_{D5}$ = 3.7 K) in (a) and (b), respectively.}
\label{fig:Dingle}
\end{figure}
%-----------------------------------------------------------------

Field dependence of the oscillatory part of the TDO frequency, obtained by removing a monotonically field-dependent background (i.e. a polynomial with a constantly positive second derivative in the studied field range), and corresponding Fourier analysis are displayed in Fig.~\ref{fig:d_TF} for two directions of the magnetic field with respect of the conducting plane ($\theta$ = 0 and 11$^{\circ}$, where $\theta$ is the angle between the field direction and the normal to the conducting $ab$ plane). Five frequencies are observed as reported in Table~\ref{tab}.

\begin{table}
\caption{Frequency ($F_i$), effective mass ($m_i^*$) and Fermi energy ($E_F$), calculated assuming parabolic dispersion, of the Fourier components with index $i$ observed in the data of Fig.~\ref{fig:d_TF}.}
\label{tab}
\begin{center}
\begin{tabular}{cccc}
$i$   & $F_i$ (T)    & $m_i^*$/$m_e$   & $E_F$ ($meV$)\\
1     & $\sim$ 50    &                 & \\
2     & 96  $\pm$ 6  & 0.75 $\pm$ 0.20 & 15 $\pm$ 5\\
3     & 200 $\pm$ 10 & 2.0 $\pm$ 0.4   & 12 $\pm$ 3\\
4     & $\sim$ 580   &                 & \\
5     & 660 $\pm$ 5  & 3.2 $\pm$ 0.6   & 24 $\pm$ 5

\end{tabular}
\end{center}
\end{table}

 These frequencies correspond to orbits area in the range 0.2-2.3 $\%$ of the first Brillouin zone area. Similar and slightly smaller frequencies are observed in magnetoresistance data of FeSe \cite{Te14} and FeTe$_{0.65}$Se$_{0.35}$\cite{Ok13}, respectively.  It should be noticed that less than 3 oscillations with the frequency $F_1$ are involved in the window field considered in Fig.~\ref{fig:d_TF} preventing any reliable data analysis, such as effective mass determination, for this Fourier component. This statement also holds for the data of Ref.~\cite{Te14}. Nevertheless, $F_1$ is close to the frequency labeled $F_{\gamma}$, observed for FeTe$_{0.65}$Se$_{0.35}$ in the field range $\sim$ 4-7 T (which involves about 4 oscillations) \cite{Ok13}. Therefore this very small frequency might be actually present in the oscillatory spectrum. Assuming circular orbits, band structure calculations predict electron and hole tubes with Fermi wave vector values corresponding to Shubnikov-de Haas frequencies of few thousands of Tesla \cite{Su08,Ku10,Ma13}. In contrast, ARPES data at low temperature \cite{Ma13} evidence only one electron and one hole orbit with Fermi wave vectors k$_F$ = 0.18 ${\AA}^{-1}$ and k$_F$ = 0.05 ${\AA}^{-1}$, respectively. Still assuming circular cross sections, these latter values correspond to frequencies of 1000 T and 80 T, respectively, which is roughly within the range of the frequency values observed in Fig.~\ref{fig:d_TF}. Low temperature ARPES data of Refs.\cite{Na14,Sh14} yield additional orbits, due to the orthorhombic distortion, with Fermi energies of few tens of a meV, in agreement with the data in Table~\ref{tab} as well. Taking FS warping into account, the number of frequencies observed should be twice the number of orbits due to necks and bellies. Besides, eventual presence of harmonics should be taken into account.  Therefore, the oscillation spectrum should be more complex than reported in Table~\ref{tab}. Nevertheless, additional frequencies such as those labeled $F'$ and $F"$ in Fig.~\ref{fig:d_TF} cannot be excluded. As for harmonics, $F_2$ and $F_3$ could be the second harmonics of $F_1$ and $F_2$, respectively. This hypothesis can be checked through the effective mass determination. Indeed, in the framework of the Lifshitz-Kosevich and Falicov-Stachowiak models \cite{Sh84,Fa66}, the effective mass $m^*_{pi}$ of the $p^{th}$ harmonics of the frequency $F_{i}$ is given by $m^*_{pi}$ = $p \times m^*_{i}$ where $m^*_{i}$ is the effective mass of the $i$ orbit. Temperature dependence of the oscillation amplitude is displayed in Fig.~\ref{fig:Dingle}a for the frequencies $F_2$, $F_3$ and $F_5$, yielding, in $m_e$ units, $m^*_2$ = 0.75 $\pm$ 0.20, $m^*_3$ = 2.0 $\pm$ 0.4 and $m^*_5$ = 3.2 $\pm$ 0.6. Owing to the error bars, it cannot be excluded that $F_3$ is the second harmonics of $F_2$. It should be noticed that such statement is at variance with previous data \cite{Te14} for which it is reported that $F_2$ is the second harmonics of $F_1$, instead, even though the effective mass relevant to $F_1$ cannot be reliably, determined as above discussed.

  Otherwise, effective mass values $m^*_2$ and $m^*_3$ are significantly smaller than those, corresponding to frequencies $F_{2\alpha}$ and $F_{\beta}$, respectively in Ref.~\cite{Te14}, measured in a lower temperature range which remains to be understood. Nevertheless, the measured values are close or even slightly larger than those deduced from quantum oscillations of underdoped cuprates with similar frequency. For example, the effective mass linked to  $F_5$ = 660  $\pm$ 5 T ($m^*_5$ = 3.2 $\pm$ 0.6) can be compared to the effective mass of YBa$_2$Cu$_3$O$_{6.5}$ ($m^*$ = 1.9 $\pm$ 0.1 for F = 530 $\pm$ 20 T) and YBa$_2$Cu$_4$O$_{8}$ ($m^*$ = 2.7 $\pm$ 0.3 for F = 660 $\pm$ 30 T), respectively \cite{Vi11}. This result is in line with the strong renormalization of the effective mass of FeSe observed by ARPES measurements \cite{Ma13,Na14,Sh14}.  Finally, the Dingle temperature deduced from the field dependence of the amplitude relevant to the frequency $F_5$ (see Fig.~\ref{fig:Dingle}b) is $T_{D5}$ = 3.7 $\pm$ 0.8 K, yielding mean free path $\lambda_5$ = 17 $\pm$ 4 nm.

\section{Summary and conclusion}

 Shubnikov-de Haas oscillations and magnetic field- and temperature-dependent superconducting transition of single crystalline FeSe have been studied by contactless tunnel diode oscillator-based measurements. In zero-field, the temperature dependence of the TDO frequency yields a superconducting transition temperature $T_c$ = 8.6 $\pm$ 0.1 K in agreement with magnetization data.

The WHH model accounts for the temperature dependence of the upper critical magnetic field for magnetic field applied both parallel (\textbf{H} $\|$ $ab$) and perpendicular to the conducting $ab$ plane (\textbf{H} $\|$ $c$). While the orbital contribution accounts for the data with \textbf{H} $\|$ $c$, a Pauli limiting contribution is evidenced for \textbf{H} $\|$ $ab$. The good agreement of the data with the WHH model suggests that superconducting properties of FeSe in magnetic field are mainly controlled by one band, only, despite the multiband nature of the Fermi surface. Finally, the anisotropy of the critical magnetic field close to $T_c$ ($\gamma$ = 4) is higher than for FeTe$_{0.5}$Se$_{0.5}$. Besides, $\gamma$ stays above 1 as the temperature goes to zero, in contrast to FeTe$_{0.5}$Se$_{0.5}$. The larger anisotropy of FeSe suggests stronger two-dimensionality.

Shubnikov-de Haas oscillations with frequencies in the range 50 T to 660 T, corresponding to orbits area from 0.2 to 2.3 $\%$ of the first Brillouin zone area have been observed in agreement with ARPES data at low temperature. Some of the effective masses,  measured in the temperature range 1.5 - 4.2 K, are significantly smaller than the recently reported values measured in a lower temperature range which remains to be understood \cite{Te14}. Nevertheless, they are at least as large as in the case of underdoped cuprates. Owing to the small frequency values, these data account for the strong band renormalization deduced from ARPES data \cite{Ma13,Na14,Sh14}.

\acknowledgments

The authors acknowledge the support of the European Magnetic Field Laboratory (EMFL) and the financial support of UJF and Grenoble INP through the AGIR-2013 contract of S.Karlsson.


\begin{thebibliography}{0}



\bibitem{Ka08}
  \Name{Y. Kamihara, T. Watanabe, M. Hirano \and H. Hosono}
  \REVIEW{J. Amer. Chem. Soc.}{130}{2008}{3296}.

\bibitem{Re08}
  \Name{Z. Ren, L. Wei, J. Yang, Y. Wei, X.-L. Shen, Z.-C. Li, G.-C. Che  X.-L. Dong, L.-L. Sun, F. Zhou \and Z.-X. Zhao}
  \REVIEW{Chin. Phys. Lett. }{25}{2008}{2215}.

\bibitem{Di08} %3
 \Name{H. Ding, P. Richard, K. Nakayama, K. Sugawara, T. Arakane,
Y. Sekiba, A. Takayama, S. Souma, T. Sato, T. Takahashi, Z. Wang, X. Dai, Z. Fang,
G. F. Chen, J. L. Luo \and N. L. Wang}
 \REVIEW{EPL}{83}{2008}{47001}.

\bibitem{Ye08} %4
 \Name{K.-W. Yeh, T.-W. Huang, Y.-l. Huang, T.-K. Chen, F.-C. Hsu, P. M. Wu, Y.-C. Lee, Y.-Y. Chu, C.-L. Chen, J.-Y. Luo, D.-C. Yan and M.-K. Wu}
 \REVIEW{EPL}{84}{2008}{37002}.

\bibitem {Ru09} %[Ru09] 5
 \Name{F. Rullier-Albenque, D. Colson, A. Forget and H. Alloul}
 \REVIEW{Phys. Rev. Lett.}{103}{2009}{057001}.

\bibitem{Bo14} %6
 \Name{A. E. B\"{o}hmer, P. Burger, F. Hardy, T. Wolf, P. Schweiss, R. Fromknecht, M. Reinecker, W. Schranz \and C. Meingast}
 \REVIEW{Phys. Rev. Lett.}{112}{2014}{047001}.

\bibitem {Zh11} %[Zh11] 7
 \Name{J.-L. Zhang, L. Jiao, Y. Chen \and H. Yuan}
 \REVIEW{Front. Phys.}{6}{2011}{463}.

\bibitem {Do13} %[Do13] 8
 \Name{C. Dong, H. Wang \and M. Fang}
 \REVIEW{Chin. Phys. B}{22}{2013}{087401}.

\bibitem {Im09} %[Im09] 9
 \Name{T. Imai, K. Ahilan, F. L. Ning, T. M. McQueen \and R. J. Cava}
 \REVIEW{Phys. Rev. Lett.}{102}{2009}{177005}.

\bibitem {Su08} % [Su08] 10
 \Name{A. Subedi, L. Zhang D. J. Singh \and M. H. Du}
 \REVIEW{Phys. Rev. B}{78}{2008}{134514}.

\bibitem {Ca11} % [Ca11] 11
 \Name{A. Carrington}
 \REVIEW{Rep. Prog. Phys.}{74}{2011}{124507}.


\bibitem {Ku10} % [Ku10] 12
 \Name{R. S. Kumar, Y. Zhang, S. Sinogeikin, Y. Xiao, S. Kumar, P. Chow, A. L. Cornelius \and C. Chen}
 \REVIEW{J. Phys. Chem. B}{114}{2010} {12597}.

\bibitem {Ma13} %[Ma13] 13
 \Name{J. Maletz, V. B. Zabolotnyy, D. V. Evtushinsky, S. Thirupathaiah, A. U. B. Wolter, L. Harnagea, A. N. Yaresko, A. N. Vasiliev,
 D. A. Chareev, E. D. L. Rienks, B. B\"{u}chner and S. V. Borisenko}
 \REVIEW{Phys.Rev.B}{89}{2014}{220506(R)}.

\bibitem {Te14} %14
 \Name{T. Terashima, N. Kikugawa, A. Kiswandhi, E.-S. Choi, J. S. Brooks, S. Kasahara, T. Watashige, H. Ikeda, T. Shibauchi, Y. Matsuda,
 T. Wolf, A. E. B\"{o}hmer, F. Hardy, C. Meingast, H. V. L\"{o}hneysen and S. Uji}
 \REVIEW{Phys.Rev.B}{90}{2014}{144517}.

\bibitem {Na14} %15
\Name{K. Nakayama, Y. Miyata, G. N. Phan, T. Sato, Y. Tanabe, T. Urata, K. Tanigaki and T. Takahashi} arXiv:1404.0857.

\bibitem {Sh14} %16
\Name{T. Shimojima, Y. Suzuki, T. Sonobe, A. Nakamura, M. Sakano, J. Omachi ,K. Yoshioka, M. Kuwata-Gonokami, K. Ono, H. Kumigashira, A. E. B\"{o}hmer, F. Hardy, T. Wolf, C. Meingast , H. v. L\"{o}hneysen, H. Ikeda and K . Ishizaka}
\REVIEW{Phys.Rev.B}{90}{2014}{121111(R)}.

\bibitem {Co13}
 \Name{A. I. Coldea, D. Braithwaite and A. Carrington}
 \REVIEW{C. R. Physique}{14}{2013}{94}. %[Co13]

\bibitem{Bl14}
 \Name{S. Blackburn, B. Pr\'{e}vost, M. Bartkowiak, O. Ignatchik, A. Polyakov, T. F\"{o}rster, M. C\^{o}t\'{e}, G. Seyfarth,
 C. Capan, Z. Fisk, R. G. Goodrich, I. Sheikin, H. Rosner, A. D. Bianchi and J. Wosnitza} \REVIEW{Phys.Rev.B}{89}{2014}{220505(R)}. %[Bl14] 15

\bibitem{Ok13}
 \Name{H. Okazaki, T. Yamaguchi, T. Watanabe, K. Deguchi, S. Demura, S. J. Denholme, T. Ozaki, Y. Mizuguchi, H. Takeya, T. Oguchi and Y. Takano}
  \REVIEW{EPL}{104}{2013}{37010}. % [Ok13]

\bibitem {Le12}
 \Name{H. Lei, D. Graf, R. Hu, H. Ryu, E. S. Choi, S. W. Tozer and C. Petrovic}
 \REVIEW{Phys.Rev.B}{85}{2012}{094515}. %[Le12] 17

\bibitem{Ve13}
 \Name{S. I. Vedeneev, B. A. Piot, D. K. Maude, and A. V. Sadakov}
 \REVIEW{Phys. Rev. B}{87}{2013}{134512}. % [Ve13] 18

\bibitem{Ka14} \Name{S. Karlsson, P. Strobel and P. Toulemonde} to be published. % [Ka14] 19

\bibitem{Ch13}
 \Name{D. Chareev, E. Osadchii, T. Kuzmicheva, J-Y. Lin, S. Kuzmichev, O. Volkovad and A. Vasiliev}
 \REVIEW{CrystEngComm}{15}{2013}{1989}. % [Ch13] 20


\bibitem{Bo13}
 \Name{A. E. B\"{o}hmer, F. Hardy, F. Eilers, D. Ernst, P. Adelmann, P. Schweiss, T. Wolf and C. Meingast}
 \REVIEW{Phys. Rev. B}{87}{2013}{180505(R)}. %[Bo13] 21


\bibitem{Dr10}
  \Name{Drigo L., Durantel F., Audouard A. \and Ballon G.}
  \REVIEW{Eur. Phys. J. Appl. Phys.}{52}{2010}{10401}. % 22

\bibitem{Pr11} \Name{R. Prozorov and V. G. Kogan}
 \REVIEW{Rep. Prog. Phys.}{74}{2011}{124505}. % [Pr11] 23

\bibitem{Ga11} \Name{V. A. Gasparov, L. Drigo, A. Audouard, D. L. Sun, C. T. Lin, S. L. Bud'ko, P. C. Canfield, F. Wolff-Fabris and J. Wosnitza}
 \REVIEW{JETP Letters}{93}{2011}{667}. %  [Ga11] 24

\bibitem{Au14}  \Name{A. Audouard, L. Drigo, F. Duc, X. Fabr\`{e}ges, L. Bosseaux and P. Toulemonde}
 \REVIEW{J. Phys.: Condens. Matter}{26}{2014}{185701}. %[Au14] 25

\bibitem{Br09}  \Name{D. Braithwaite, B. Salce, G. Lapertot, F. Bourdarot, C. Marin, D. Aoki and M. Hanfland}
 \REVIEW{J. Phys.: Condens. Matter}{21}{20009}{232202}.

\bibitem{Ye13}  \Name{S. Yeninas, M. A. Tanatar, J. Murphy, C. P. Strehlow, O. E. Ayala-Valenzuela, R. D. McDonald, U. Welp,
 W. K. Kwok, T. Kobayashi, S. Miyasaka, S. Tajima, and R. Prozorov}
 \REVIEW{Phys. Rev. B}{87}{2013}{094503}. %[Ye13] 26

\bibitem{WHH} \Name{N. R. Werthamer, K. Helfand and P. C. Hohnenberg}
 \REVIEW{Phys. Rev.}{147}{1966}{295}. % [WHH] 27

\bibitem {Kl10}
 \Name{T. Klein, D. Braithwaite, A. Demuer, W. Knafo, G. Lapertot, C. Marcenat, P. Rodi\`{e}re, I. Sheikin, P. Strobel, A. Sulpice and P. Toulemonde}
 \REVIEW{Phys. Rev. B}{82}{2010}{184506}. %[Kl10] 28

\bibitem{Cl62} \Name{A. M. Clogston}
 \REVIEW{Phys. Rev. Lett.}{9}{1962}{266}. % [Cl62] 29

\bibitem {Kh10} \Name{R. Khasanov, M. Bendele, A. Amato, K. Conder, H. Keller, H.-H. Klauss, H. Luetkens and E. Pomjakushina}
 \REVIEW{Phys. Rev. Lett.}{104}{2010}{087004}.%[Kh10] 30

\bibitem{Li11} \Name{J.-Y. Lin, Y. S. Hsieh, D. A. Chareev, A. N. Vasiliev, Y. Parsons and H. D. Yang}
  \REVIEW{Phys. Rev. B}{84}{2011}{220507(R)}.% [Li11] 31

\bibitem {Se10} \Name{A. Serafin, A. I. Coldea, A. Y. Ganin, M. J. Rosseinsky, K. Prassides, D. Vignolles and A. Carrington}
 \REVIEW{Phys. Rev. B}{82}{2010}{104514}.% [Se10] 32

\bibitem {Br10} \Name{D. Braithwaite, G. Lapertot, W. Knafo and I. Sheikin}
 \REVIEW{J. Phys. Soc. Jap.}{79}{2010}{053703}.%[Br10] 33

\bibitem{Gu03} \Name{A. Gurevich}
 \REVIEW{Phys. Rev. B}{67}{2003}{184515}. %  [Gu03] 34

\bibitem{Le10}  \Name{H. C. Lei, R. W. Hu, E. S. Choi, J. B. Warren and C. Petrovic}
  \REVIEW{Phys. Rev. B}{81}{2010}{94518}. % [Le10] 35

\bibitem{Sh84}
  \Name{Shoenberg D.}
  \Book{Magnetic Oscillations in Metals}
  \Editor{Cambridge University Press}
  %\Vol{}
  \Publ{Cambridge, England}
  \Year{1984}
  %\Page{}.

\bibitem{Fa66} \Name{L. M. Falicov and H. Stachowiak}
  \REVIEW{Phys. Rev.}{147}{1966}{505}.% [Fa66] 37

\bibitem{Vi11} \Name{B. Vignolle, D. Vignolles, D. LeBoeuf, S. Lepault, B. Ramshaw, R. Liang, D. A. Bonn, W.N. Hardy, N. Doiron-Leyraud,
 A. Carrington, N. E. Hussey, L. Taillefer and C. Proust}
\REVIEW{C. R. Physique}{12}{2011}{446}. % [Vi11]  38

\end{thebibliography}
\end{document}